\documentclass[10pt,conference,letterpaper]{IEEEtran}
\makeatletter
\def\ps@headings{%
\def\@oddhead{\mbox{}\scriptsize\rightmark \hfil \thepage}%
\def\@evenhead{\scriptsize\thepage \hfil \leftmark\mbox{}}%
\def\@oddfoot{}%
\def\@evenfoot{}}
\makeatother
\usepackage{richard_config}

\begin{document} 

\title{Coordination of autonomic functionalities in communications networks}

\author{
\IEEEauthorblockN{Richard Combes\IEEEauthorrefmark{1}, Zwi Altman\IEEEauthorrefmark{1} and Eitan Altman\IEEEauthorrefmark{2}}

\IEEEauthorblockA{\IEEEauthorrefmark{1}Orange Labs\\
38/40 rue du G\'en\'eral Leclerc,92794 Issy-les-Moulineaux\\
Email:\{richard.combes,zwi.altman\}@orange.com}

\IEEEauthorblockA{\IEEEauthorrefmark{2}INRIA Sophia Antipolis\\
06902 Sophia Antipolis, France\\
Email:Eitan.Altman@sophia.inria.fr}
}

\date{\today}
\maketitle

\begin{abstract}
	Future communication networks are expected to feature autonomic (or self-organizing) mechanisms to ease deployment (self-configuration), tune parameters automatically (self-optimization) and repair the network (self-healing). Self-organizing mechanisms have been designed as stand-alone entities, even though multiple mechanisms will run in parallel in operational networks. An efficient coordination mechanism will be the major enabler for large scale deployment of self-organizing networks. We model self-organizing mechanisms as control loops, and study the conditions for stability when running control loops in parallel. Based on control theory and Lyapunov stability, we propose a coordination mechanism to stabilize the system, which can be implemented in a distributed fashion. The mechanism remains valid in the presence of measurement noise via stochastic approximation. Instability and coordination in the context of wireless networks are illustrated with two examples and the influence of network geometry is investigated. We are essentially concerned with linear systems, and the applicability of our results for non-linear systems is discussed.
\end{abstract}
\section{Introduction}
Deployment, optimization and maintenance of communication networks are complex tasks which occupy thousands of network engineers everyday. In order to ease this burden and reduce costs of operations, researchers and industrials have proposed to include autonomic functionalities in future networks. Networks with autonomic entities are often called \ac{SON}. Self-organization should enable at least partial automation of the configuration of newly deployed network node (self-configuration), of parameter tuning (self-optimization) and of reparation of faulty network nodes (self-healing).

In wireless networks, SON functionalities such as \ac{ICIC}, load balancing, management of \ac{BS} sleep mode, mobility management and drive test automation have been identified as important use cases (\cite{3gpp.36.902}). Future standards for wireless networks such as \ac{LTE} feature \ac{SON} functionalities. In previous contributions, \ac{SON} functionalities have been designed as stand-alone entities (\cite{StolyarInfocom2009,CombesPeva2011,CombesInfocom2012,ElayoubiSakerinfocom2011,CombesCNSM2011}). Such contributions did not take into account the interaction between different \ac{SON} functionalities operating simultaneously. Since \ac{SON} functionalities are numerous, a large number of \ac{SON} entities might operate simultaneously in the network. This poses a fundamental stability issue, if we cannot guarantee stable interaction of multiple \ac{SON} entities. It is fair to say that a robust coordination mechanism will be the main enabler for large-scale deployment of \ac{SON}.

\emph{Our contribution:} The main contribution of this paper is to propose a generic mathematical model for the interaction of multiple \ac{SON} mechanisms running in parallel, along with a practically implementable coordination mechanism. We model \ac{SON} mechanisms as control loops, so that the system can be described by an \ac{ODE}. Stability is studied using the Lyapunov approach used by control theorists. In particular, stability conditions can be stated in terms of \emph{linear matrix inequalities} (\cite{LMIControlBoyd}). When stability does not hold naturally, we propose a \emph{coordination} mechanism which forces stability. The design of a coordination mechanism corresponds, in the context of control theory, to the concepts of \emph{controllability} and \emph{state-feedback synthesis}. We propose a generic coordination mechanism which can be implemented in a distributed way. When the network is controlled using measurements corrupted by additive noise, the results remain valid using stochastic approximation theorems (\cite{Kushner}). 

	To the best of our knowledge, no previous contribution had studied the problem of \ac{SON} coordination using this control theory/stochastic approximation-based framework and provided a generic coordination mechanism which is practically implementable (distributed and possibly asynchronous).

The rest of the paper is organized as follows: in Section~\ref{sec:parallel_control_loops} we state the proposed model for interaction of \ac{SON} mechanisms running in parallel and the coordination problem to be solved. In Section~\ref{sec:linear_case} we examine the case where performance indicators are linear functions of the parameters, and propose a practically implementable coordination mechanism. In Section~\ref{sec:application} we illustrate the application of our model to traffic management in wireless networks with two examples. Section~\ref{sec:conclusion} concludes the paper. In appendices~\ref{app:odes} and~\ref{app:linear_ode} we recall the basic notions of stability for \acp{ODE} and linear \acp{ODE} respectively.
\section{SON coordination as parallel control loops}\label{sec:parallel_control_loops}
\subsection{The model}
	A \ac{SON} mechanism is an entity which monitors a given performance indicator and controls a scalar parameter. The current value of the performance indicator is observed, and the parameter is modified accordingly to attain some objective. We consider $I > 1$ \ac{SON} mechanisms operating simultaneously. We define $\theta_i$ the parameter controlled by the $i$-th \ac{SON} mechanism and $\theta = (\theta_1,\dotsc,\theta_I)$ the vector of parameters. The $i$-th \ac{SON} mechanism monitors a performance indicator and updates its parameter $\theta_i$ proportionally to $F_i(\theta)$. $F(\theta) = (F_1(\theta),\dotsc,F_I(\theta))$ is the direction of update of $\theta$.

	We say that the $i$-th \ac{SON} mechanism operates in \emph{stand-alone} mode if all parameters but $\theta_i$ are kept constant. Namely all the other mechanisms are shut down. The $i$-th \ac{SON} mechanism operating in stand-alone is described by the \ac{ODE}:
\begin{align}\label{eq:son_standalone}
	\dot{ \theta_i } &= F_i(\theta), \nonumber\\
	\dot{ \theta_j } &= 0 , j \neq i.
\end{align}
	We say that the \ac{SON} mechanisms operate in \emph{parallel} mode if all parameters are modified simultaneously, which is described by the \ac{ODE}:
\begin{align}\label{eq:son_simultaneous}
	\dot{ \theta } &= F(\theta).
\end{align}
 	We say that the $i$-th \ac{SON} mechanism is stable in stand-alone mode if there exists $\theta^{*,i}_i$ which is \emph{asymptotically stable} for~\eqref{eq:son_standalone}. The definition of asymptotic stability is recalled in appendix~\ref{app:odes}. It is noted that $\theta^{*,i}_i$ can depend on $\theta_j$, $j \neq i$. We say that the \ac{SON} mechanisms are stable in parallel mode if there exists $\theta^*$ which is asymptotically stable for~\eqref{eq:son_simultaneous}. Typically, the \ac{SON} mechanisms are designed and studied in a stand-alone manner, so that stand-alone stability is verified. 
	
	However, stand-alone stability does not imply parallel stability. First consider a case where $F_i(\theta)$ does not depend on $\theta_j$, for all pairs $(i,j)$ , $j \neq i$. Then~\eqref{eq:son_simultaneous} is a set of $I$ parallel independent \acp{ODE}, so that stand-alone stability implies parallel stability. On the other hand, if there exists $i \neq j$ such that $F_i(\theta)$ depends on $\theta_j$, then the situation is not so clear-cut. We say that \ac{SON} $i$ and $j$ \emph{interact}. Namely, interaction potentially introduces \emph{instablity}.
	
	In the remainder of this article we will be concerned with conditions for parallel stability, and designing coordination mechanisms to force stability whenever possible.
\subsection{Examples of parallel mechanisms}
	Two particular cases of parallel mechanisms will be of interest. The first case is what we will call \emph{zero-finding} algorithms. Each \ac{SON} mechanism monitors the value of a performance indicator and tries to achieve a fixed target value for this performance indicator. Namely:
\begin{equation}\label{eq:son_target}
	F_i(\theta) =  f_i(\theta) - \overline{f}_i,
\end{equation}
	where $f_i$ is the performance indicator monitored by \ac{SON} $i$ and $\overline{f}_i$ - a target level for this performance indicator. The goal of the $i$-th \ac{SON} mechanism is to find $\theta_i^*$ such that $f_i(\theta^*) = \overline{f}_i$. If $\theta_i \mapsto f_i(\theta_1,\dotsc,\theta_i,\dotsc,\theta_I)$ is strictly decreasing $1 \leq i \leq I$ then stand-alone stability is assured. 

	Another case of interest is maximization algorithms. Each \ac{SON} mechanism tries to maximize a given performance indicator. There exists a continuously differentiable function $g_i$ such that:
\begin{equation}\label{eq:son_maximize}
	F_i(\theta) =  \nabla_{\theta_i} g_i(\theta).
\end{equation}
	In stand-alone mode, \ac{SON} $i$ indeed converges to a local maximum of $g_i$. If we restrict $\theta$ to a closed, convex and bounded set and if $\theta_i \mapsto g_i(\theta_1,\dotsc,\theta_i,\dotsc,\theta_I)$ is concave $1 \leq i \leq I$ , we fall within the framework of \emph{concave games} considered in \cite{RosenConvexGame}. An important result of \cite{RosenConvexGame} is that if we add an assumption called \emph{diagonal strict convexity}, then parallel stability occurs. However, diagonal strict convexity is fairly restrictive, and without it there is no guarantee that parallel stability occurs, and coordination is needed.
\subsection{Discrete time algorithms}
	Our model is based on \acp{ODE}, which are both \emph{deterministic} and \emph{continuous-time} objects. In a practical system, the parameters evolve in discrete time. Furthermore, the parameters evolve stochastically because they are updated based on noisy feedback, due to measurement noise. Denote by $\theta[t] = (\theta_1[t], \dotsc, \theta_I[t])$ the value of parameters at time $t \in \NN$. At each parameter update, performance indicators are estimated based on measurements so that \ac{SON} $i$ obtains the quantity $F_i(\theta[t]) + M_i[t]$, with $\expec{M_i[t]} = 0$. The additive noise $\Set{M[t]}_{t \in \NN}$ is introduced by the measurements. The parameters are updated using the noisy feedback:
\begin{equation}\label{eq:son_stoch_approx}
	\theta[t+1] = \theta[t] + \epsilon (F(\theta[t]) + M[t]).
\end{equation}
where $\epsilon > 0$ is a small constant.	Stochastic approximation (\cite{Kushner,Borkar}) gives a strong link between the noisy, discrete-time~\eqref{eq:son_stoch_approx} and the \ac{ODE} studied in our model~\eqref{eq:son_simultaneous}. In particular it can be shown that (under some technical conditions) $\Set{\theta[t]}_{t \in \NN}$ converges to \emph{asymptotically stable} sets of the \ac{ODE} when $\epsilon \to 0^+$. The main point is that our model based on \acp{ODE} is sufficient to study the stability of the discrete-time, noisy algorithms used in practical systems. 

\section{Coordination for linear systems}\label{sec:linear_case}
\subsection{Linear systems}
		In most of the remainder of this paper, we will study the case where $F$ is affine:
\begin{equation}\label{eq:son_linear}
	F(\theta) = A \theta + b,
\end{equation}
with $b$ a vector of size $I$ and $A$ a matrix of size $I \times I$. We assume that $A$ is invertible and we define $\theta^* = -A^{-1} b$. 
	The \ac{SON} mechanisms running in parallel are described by the linear \ac{ODE}:
\begin{equation}\label{eq:son_linear_ode}
	\dot{\theta} = A \theta + b = A (\theta - \theta^*).
\end{equation}
	It is noted that in the linear case, we always fall within the scope of zero-finding algorithms described previously, by defining:
\begin{align}\label{eq:son_linear_ode_zero}
	f_i(\theta) &=  \sum_{1 \leq j \leq I} A_{i,j} \theta_j, \\
	\overline{f}_i &= -b_i, \\
	\dot{\theta}_i &= f_i(\theta) - \overline{f}_i.
\end{align}
	In particular, stand-alone stability occurs iif $A_{i,i} < 0$ , $1 \leq i \leq I$, i.e all the diagonal terms of $A$ are strictly negative. Basic results on linear \acp{ODE} are recalled in appendix~\ref{app:linear_ode}. Namely, parallel stability holds iif all the eigenvalues of $A$ have strictly negative real part.

\subsection{Coordination}
\subsubsection{Coordination mechanism}
	If $A$ has at least one eigenvalue with positive or null real part, convergence to $\theta^*$ does not occur, and a coordination mechanism is needed. We consider a \emph{linear} coordination mechanism, where $A$ is replaced by $C A$ with $C$ a $I \times I$ real matrix. The \ac{ODE} for the coordinated system is:
\begin{equation}\label{eq:son_linear_ode_coord}
	\dot{\theta} = C A (\theta - \theta^*).
\end{equation}
	 Define $c_i$ as:
\begin{equation}\label{eq:son_coord_kpi}
	c_i(\theta) =  \sum_{1 \leq j \leq I} C_{i,j} f_j(\theta).
\end{equation}
	The coordination mechanism can be interpreted as transforming the performance indicator monitored by \ac{SON} $i$ from $f_i$ to a linear combination of the performance indicators monitored by all the \ac{SON} mechanisms $c_i$. As explained in appendix~\ref{app:linear_ode}, stability is achieved if there exists a symmetric matrix $X$  such that:
\begin{align}
	(C  A)^T X + X C A \prec 0,\\
	0 \prec X,
\end{align}
where $\prec$ denotes positive definitiveness for symmetric matrices. In particular, 
\begin{equation} 
V(\theta) = (\theta - \theta^*)^T X (\theta - \theta^*),
\end{equation} 
acts as a Lyapunov function.
\subsubsection{Distributed implementation}	
	It is noted that the choice for the coordination matrix $C$ is not unique. For instance $C = A^{-1}$ ensures stability. For the coordination mechanism to be scalable with respect to the number of \acp{SON}, $C$ should be chosen to allow \emph{distributed} implementation. We say that \ac{SON} $j$ is a neighbor of \ac{SON} $i$ if $\frac{\partial f_j}{\partial \theta_i} \neq 0$. We define ${\cal I}_i$ the set of neighbors of $i$. The coordination mechanism is distributed if each \ac{SON} needs only to exchange information with its neighbors.
	
	We give an example of a coordination mechanism which can always be distributed. The mechanism is based on a \emph{separable} Lyapunov function. Define the weighted square error:
\begin{equation}
	V(\theta) = \sum_{i=1}^I w_i (f_i(\theta) - \overline{f}_i)^2 = (\theta - \theta^*)^T A^T W A (\theta - \theta^*),
\end{equation}
with $\Set{w_i}_{1 \leq i \leq I}$ strictly positive weights and $W$ the diagonal matrix with diagonal elements $\Set{w_i}_{1 \leq i \leq I}$.
	Coordination is achieved by following the gradient of $-V$ so that $V$ is a Lyapunov function:
\begin{equation}
	\dot{\theta} = - \nabla_{\theta} V(\theta) =  - A^T W A (\theta - \theta^*).
\end{equation}
	Namely, we choose $C = - A^T W$. We can verify that the mechanism is distributed:
\begin{equation}
	\dot{\theta}_i =  \sum_{j=1}^I  2 w_i  \frac{\partial f_j}{\partial \theta_i}  (  f_j(\theta) - \overline{f}_j ) =  \sum_{j \in {\cal I}_i} 2 w_j  \frac{\partial  f_j}{\partial  \theta_i}  (  f_j(\theta) - \overline{f}_j ).
\end{equation}
	Indeed, the update of $\theta_i$ only requires knowledge of $\frac{\partial f_j}{\partial \theta_i}$ and $f_j(\theta) - \overline{f}_j$, for $j \in {\cal I}_i$, and this information is available from the \emph{neighbors} of $i$.
	
	It is also noted that such a mechanism can be implemented in an \emph{asynchronous manner}, i.e the components of $\theta$ are updated in a round-robin fashion, or at random instants, and the average frequency of update is the same for all components. The reader can refer to \cite{Bertsekas}[chapters 6-8] for the round-robin updates  and \cite{Kushner}[chapter 12] for the random updates. Asynchronous implementation is important in practice because if the \acp{SON} are not co-located, maintaining clock synchronization among the \acp{SON} would generate a considerable amount of overhead.
\subsection{Applicability of the linear model and parameter estimation}
	For practical systems, performance indicators $F(\theta)$ are not linear functions of $\theta$. However, as long as they are smooth, they can be approximated by linear functions using a Taylor expansion. Consider a point $\theta^*$ with $F(\theta^*) = 0$. If the values of $\theta$ are restricted to a small neigborhood of $\theta^*$:
\begin{equation} 
	F(\theta) \approx J F(\theta^*) (\theta - \theta^*),
\end{equation}
with $J F(\theta^*)$ the Jacobian of $F$ evaluated at $\theta^*$. The Hartman-Grossman theorem (\cite{HartmanGrossman}) states that on a neigbourhood of $\theta^*$, stability of the system with linear approximation implies stability of the original, non-linear system. Hence implementing the proposed coordination mechanism where $A$ is replaced by $J F$ ensures stability if we constrain $\theta$ to a small neighborhood of $\theta^*$.
	
	The parameters $A$ and $b$ might be unknown, and we can only access to noisy values of $F(\theta)$ for different values of $\theta$. The crudest approach is to estimate $A$ and $b$ through finite differences:
	\begin{align}
	a_{i,j} &\approx \frac{ f_j(\theta +  e_i \delta \theta_i) - f_j(\theta -  e_i \delta \theta_i)  }{ 2 \delta \theta_i }, \\
	b_{i} &\approx f_i(0).
	\end{align}
	with $e_i$ the $i$-th unit vector. The results are averaged over several successive measurements and additive noise is omitted for notation clarity. In general, the measurements of $F$ are obtained by calculating the time average of some output of the network during a relatively long time, so that a form of the central limit theorem applies and the additive noise is Gaussian. In this case, a better method is to employ \emph{least-squares regression}. Least-squares regression is a well studied topic  with very efficient numerical methods (\cite{LeastSquaresBjorck}) even for large data sets so that estimation of $A$ and $b$ is not computationally difficult.
	
	Finally, since practical systems do not remain stationary for an infinite amount of time, a database with values of $A$ and $b$ for each set of operating conditions must be maintained. In the context of wireless networks, the relationship between parameters and performance indicators changes when the traffic intensity changes because of daily traffic patterns. For instance, during the night traffic is very low, and traffic peaks are observed at the end of the day. A database with estimated values of $A$ and $b$ at (for instance) each hour of the day should be constructed. 
	
\section{Application to wireless networks}\label{sec:application}
	In this section we illustrate instability and coordination in the context of wireless networks using two examples. We show that instability occurs even in simple models with as few as two \acp{SON} in parallel.
\subsection{Admission control and resource allocation}
\subsubsection{Model}
	We consider a \ac{BS} in downlink, serving elastic traffic. Users enter the network according to a Poisson process with arrival rate $\lambda$, to download a file of exponential size $\sigma$, with $\expec{\sigma} < +\infty$. The \ac{BS} has $x_{max}$ parallel resources available, and we write $x \in [0,x_{max}]$ the amount of resources used. We ignore the granularity of resources, either assuming that there are a large number of resources or using time sharing, using each resource a proportion $\frac{x}{x_{max}}$ of the time. Depending on the access technology, resources can be: codes in \ac{CDMA}, time slots in \ac{TDMA}, time-frequency blocks in \ac{OFDMA} etc. When a user is alone in the system, his data rate is $R x$. Users are served in a processor sharing manner (for instance through Round Robin scheduling): if there are $n$ active users, each user has a throughput of $\frac{x R}{n}$. Admission control applies. We define $b \geq 0$ a blocking threshold and the probability of accepting a new user when $n$ users are already present in the system is $\phi(n - b)$ with $\phi:\RR \to [0,1]$ a smooth, strictly decreasing function and $\phi(n) \tends{n}{+\infty} 0$. We choose $\phi$ as a logistic function for numerical calculations:
\begin{equation} 
	\phi(n) = \frac{1}{1 + e^{n}}.
\end{equation}
	
	Define $n(t)$ the number of active users in the system at time $t$, then $n(t)$ is a continuous time Markov chain. $n(t)$ is ergodic because the probability of accepting a new user goes to $0$ as $n \to \infty$. We define the load:
\begin{equation}
	\rho(x) = \frac{\lambda \expec{\sigma} }{x R}.
	\label{eq:load_defintion}
\end{equation}
We write $\pi$ the stationary distribution of $n(t)$. $n(t)$ is reversible, and $\pi$ can be derived from the detailed balance conditions:
\begin{equation}
	\pi(n,x,b) = \frac{ \rho(x)^n \prod_{k=0}^{n-1} \phi(k-b) }{\sum_{n \geq 0}  \rho(x)^n \prod_{k=0}^{n-1} \phi(k-b) }.
	\label{eq:statonary_distribution}
\end{equation}
Using Little's law, the mean file transfer time is given by the average number of active users divided by the arrival rate:
\begin{equation}
	T(x,b) =\frac{1}{\lambda} \sum_{n \geq 0} n\pi(n,x,b).
	\label{eq:file_transfer}
\end{equation}
Let $R_{min}$ a minimal data rate required to ensure good \ac{QoS}. We say that there is an outage in a state of the system if users have a throughput lower than $R_{min}$. When there are $n$ active users in the system, outage occurs if and only if:
\begin{equation}
	n > \frac{xR}{R_{min}}.
	\label{eq:outage_cond}
\end{equation}
The outage probability is then:
\begin{equation}
	O(x,b) = \sum_{n \geq 0} \pi(n,x,b) \indic_{ (0,+\infty) }\left(n - \frac{xR}{R_{min}}  \right).
	\label{eq:outage}
\end{equation}
In this model, $x \to O(x,b)$ is not smooth, which is why we introduce the smoothed outage $\tilde{O}$: 
\begin{equation}
	\tilde{O}(x,b) = \sum_{n \geq 0} \pi(n,x,b)  \psi \left(n - \frac{xR}{R_{min}}\right).
	\label{eq:outage_smooth}
\end{equation}
with $\psi$ a smooth function approximating  $\indic_{(0,+\infty)}$. We also choose $\psi$ as a logistic function for numerical calculations.

 This queuing system is controlled by two mechanisms, and that control occurs on a time scale much slower than the arrivals and departures of the users, so that the mean file transfer time and outage probability are relevant performance metrics, and can be estimated from (noisy) measurements. The mechanisms are:
\begin{itemize}
	\item \emph{Resource allocation:} a mechanism adjusts the amount of used resources to reach a target outage rate. Such mechanisms have been considered in green networking when a \ac{BS} can switch off part of its resources in order to save energy. 
		
	Another application is interference coordination: using more resources will create inter-cell interference in neighboring \acp{BS} and degrade their \ac{QoS}. Hence \acp{BS} should use as little resources as possible, as long as their target \ac{QoS} is met.
	\item \emph{Admission control: } another mechanism adjusts the admission control threshold to reach a target file transfer time. In particular, it is noted that without admission control, the mean file transfer time becomes infinite in overload.
\end{itemize}
It is noted that $x \to \tilde{O}(x,b)$ is strictly decreasing and \newline $b \to T(x,b)$ is strictly increasing. Using the notations of Section \ref{sec:parallel_control_loops}, we have $I = 2$ control loops, $\theta_1 \equiv x$, $\theta_2 \equiv b$, $f_1 \equiv \tilde{O}$, $f_2 \equiv -T$. Consider $\theta^* = (x^*,b^*)$ an operating point. The stability in the neighborhood of $(x^*,b^*)$ can be calculated. The system will fail to converge to the desired operating point as long as the Jacobian matrix has a negative determinant, so that there are two eigenvalues of opposite sign:
	\begin{equation}
		-\frac{\partial \tilde{O} }{\partial x} \frac{\partial T}{\partial b}(x^*,b^*) +  \frac{\partial \tilde{O} }{\partial b } \frac{\partial T}{\partial x}(x^*,b^*) < 0
		\label{eq:instability_condition}
	\end{equation}
\subsubsection{Results}
\begin{figure}[htbp]
		\begin{center}
			\includegraphics[width=\columnwidth]{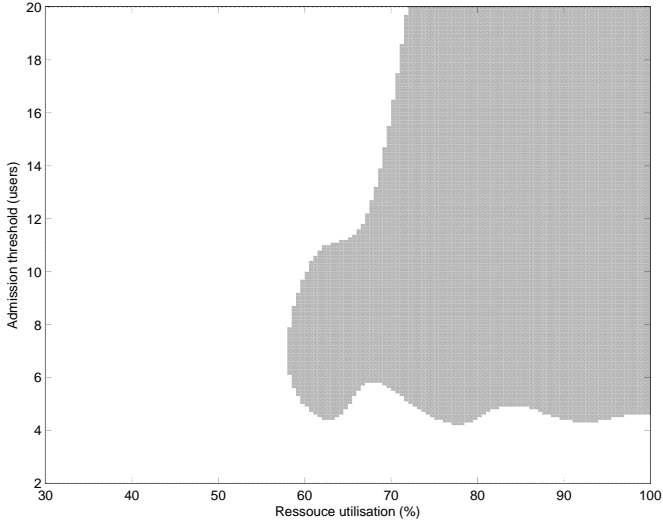}
		\end{center}
		\caption{Stability region of the system}
		\label{fig:stability_region}
	\end{figure}

	We now evaluate the stability region of the system numerically by checking condition~\eqref{eq:instability_condition} for various operating points. We choose the following parameter values: $\lambda = 0.5 users/s$, $\expec{\sigma} = 10 Mbits$, $R = 15 Mbits/s$, $R_{min} = 2 Mbits/s$, $x_{max} = 1$. Figure~\ref{fig:stability_region} presents the results. In the white region the system is stable, and in the gray region it is unstable. Even in such a simple setting with $1$ \ac{BS} and $2$ \ac{SON} mechanisms, instability occurs for a large set of operating points.
\subsection{Distributed interference control}
\subsubsection{Model}
	We turn to a case in which the \acp{SON} are not co-located, and distributed coordination is required. We study a model in which each \ac{BS} can choose its transmitted power in order to control the \ac{QoS} in the network. We consider a dense network, and the objective of each \ac{BS} is to make sure that their neighboring \acp{BS} achieve a target coverage probability, by not transmitting at too high power. We define the coverage probability as the proportion of users whom achieve a given minimum data rate. This is relevant to current wireless networks, where \acp{BS} are linked with an interface (called X2 interface in the \ac{LTE} standard), so that a \ac{BS} experiencing low data rates or congestion can send an alarm its neighbors to force them to transmit at lower power to reduce inter-cell interference, or to offload some of its traffic to reduce congestion.
	
	We consider $N_s$ \acp{BS} serving a bounded area $\AAb$. The zone served by $\ac{BS}$ $i$ is written $\AAb_i$. Interference from neighboring cells is treated as Gaussian noise. We denote by $h_i(r)$ the signal attenuation between \ac{BS}  $i$ and location $r$. $h_i(r)$ includes both path-loss and shadowing. For numerical calculations, we use the classical model where the signal attenuation (in dB) is a linear function of the distance, and shadowing is (in dB) a centered Gaussian random variable. Fast-fading is ignored. \ac{BS} $i$ transmits at power $P_i$. The \ac{SINR} at location $r \in \AAb_s$ is calculated by:
\begin{equation}
	S_i(r) = \frac{ h_i(r) P_i }{ N_0 + \sum_{j \neq i} h_{j}(r) P_{j}},
\end{equation}
with $N_0$ the thermal noise.
	We denote by $R_i(r)$ the data rate at location $r$ when no other users are being served by \ac{BS} $i$. The data rate $R_i(r)$ is calculated using the Shannon formula:
\begin{equation}
	R_i(r) =  w \log_2(1 + S_i(r)),
\end{equation}
with $w$ the system bandwidth. Consider a target data rate $R_{min}$. We define $K_i$ the coverage probability for \ac{BS} $i$, which is the probability that a user arrives in $\AAb_i$ at a location $r$ such that $R_i(r) \geq R_{min}$. The coverage probability can be calculated by:
\begin{equation}
	K_i = \frac{1}{\abs{\AAb_i}} \int_{\AAb_i} \indic_{ \Set{ R_i(r) \geq R_{min}}  }(r)dr.
\end{equation}
	Consider users arriving in the network according to a space-time Poisson process of intensity $\lambda$, i.e the number of users arriving during time $dt$ in a region of size $dr$ centered at location $r$ is $\lambda dr dt$. Then the number of users arriving during time interval $[0 , T]$ in $\AAb_i$ which have a data rate superior to $R_{min}$, divided by $\abs{\AAb_i} \lambda T$ is an unbiased estimate of $K_i$. We define ${\cal B}_i$ the set of neighbors of \ac{BS} $i$. The neigbouring relation can either come from geographical proximity, or from propagation conditions, so that neighboring \acp{BS} are \acp{BS} which strongly interfere each other.
 	
 We define $G_i$ the coverage probability of the neighbours of \ac{BS} $i$ by:
 \begin{equation}
 G_i =  \frac{\sum_{j \in {\cal B}_i } \abs{\AAb_j} F_j}{\sum_{j \in {\cal B}_i } \abs{\AAb_j}}.  
 \end{equation}
	We study the case where each \ac{BS} adjusts its transmitted power $P_i$ to avoid degrading the network \ac{QoS} and make sure that its neighbors reach a target coverage probability $\overline{G}_i$. \ac{BS} $i$ transmitting at higher power decreases the data rate of users in $\AAb_j$ , $j \in {\cal B}_i$, so that $P_i \mapsto G_i$ is strictly decreasing. It is noted that we work with $P_i$ in dB (not in linear scale). Using the notations of Section \ref{sec:parallel_control_loops}, we have $I = N_s$ \acp{SON}, with $\theta_i \equiv P_i$, and $f_i \equiv G_i$.
	
	\subsubsection{Results}
	For numerical calculations, the signal attenuation (in dB) at distance $d$ (in km) is $128 + 36.4 \log_{10}(d)$. The shadowing standard deviation is $6dB$. Thermal noise power spectral density is $-174dBm/Hz $, the system bandwidth $w$ is $20 MHz$. The minimal data rate $R_{min}$ is $20Mbits/s$, corresponding to a minimal \ac{SINR} of $0dB$. 
	
\begin{figure}[htbp]
		\begin{center}
			\includegraphics[width=\columnwidth]{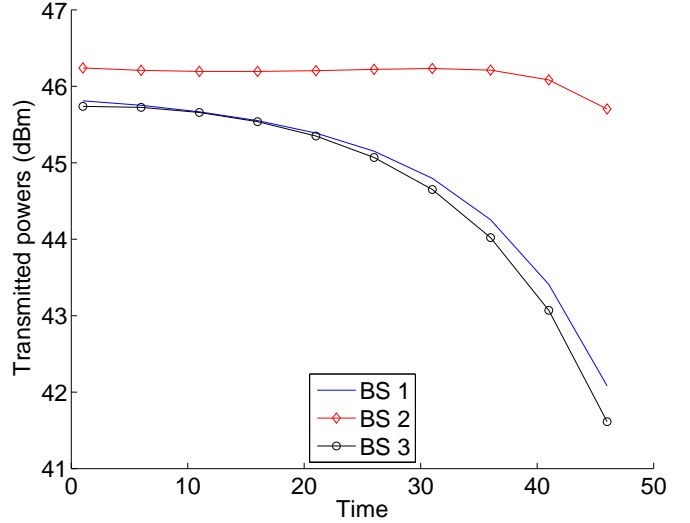}
		\end{center}
		\caption{Hexagonal network, no coordination, transmitted powers as a function of time}
		\label{fig:hexa_network_powers}
	\end{figure}

\begin{figure}[htbp]
		\begin{center}
			\includegraphics[width=\columnwidth]{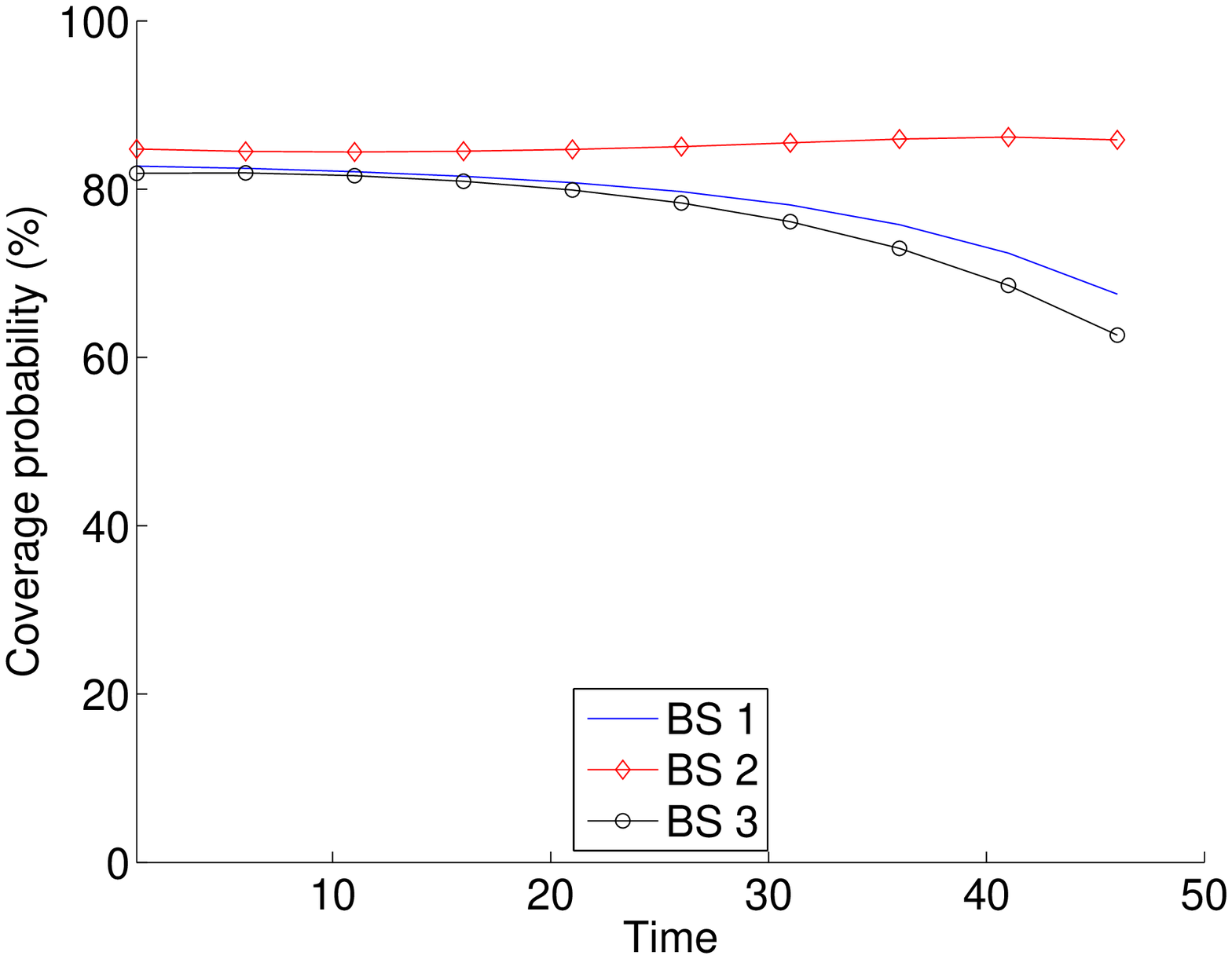}
		\end{center}
		\caption{Hexagonal network, no coordination, coverage probability as a function of time}
		\label{fig:hexa_network_coverage}
	\end{figure}

\begin{figure}[htbp]
		\begin{center}
			\includegraphics[width=\columnwidth]{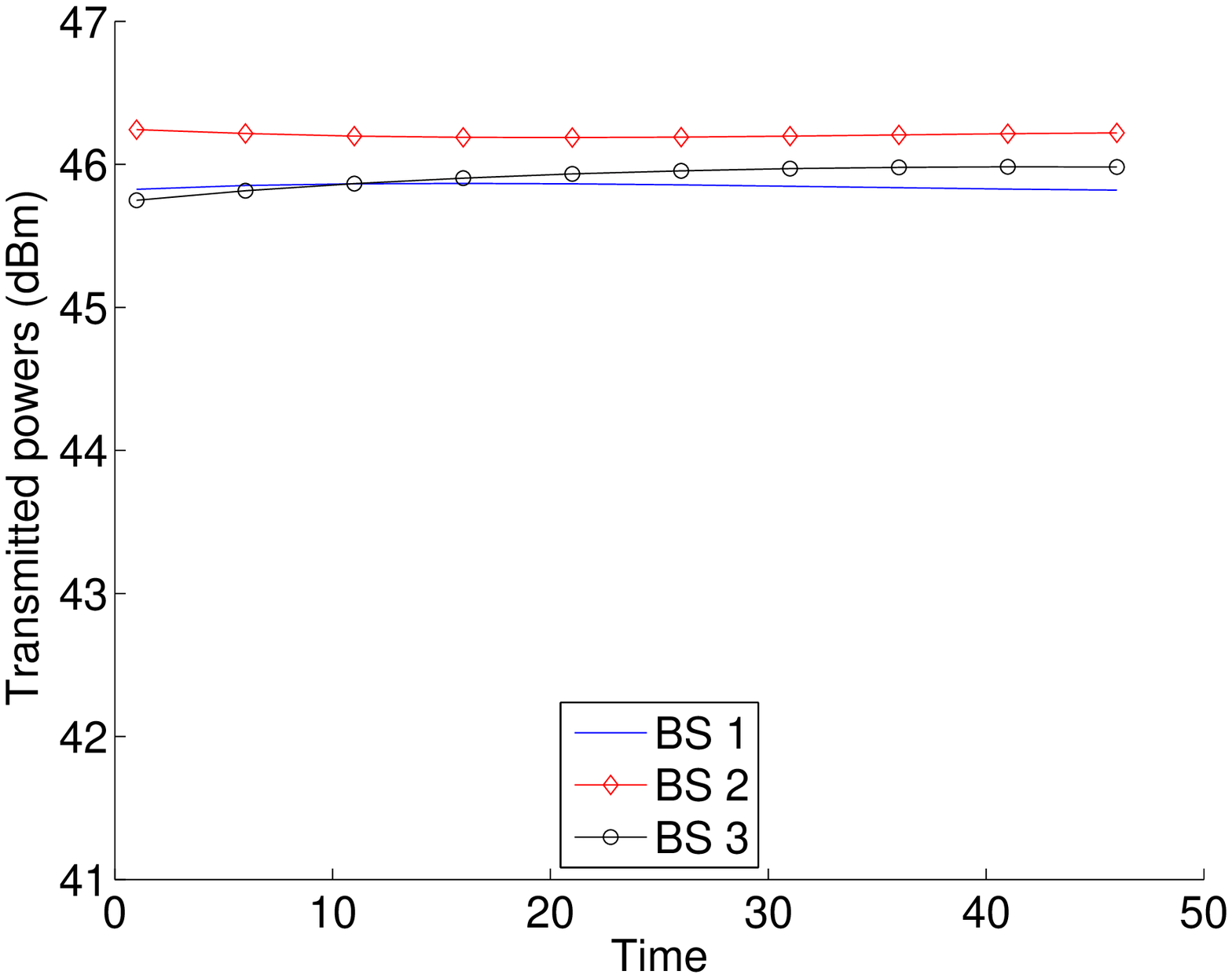}
		\end{center}
		\caption{Hexagonal network with coordination, transmitted powers as a function of time}
		\label{fig:hexa_network_powers_coord}
	\end{figure}

\begin{figure}[htbp]
		\begin{center}
			\includegraphics[width=\columnwidth]{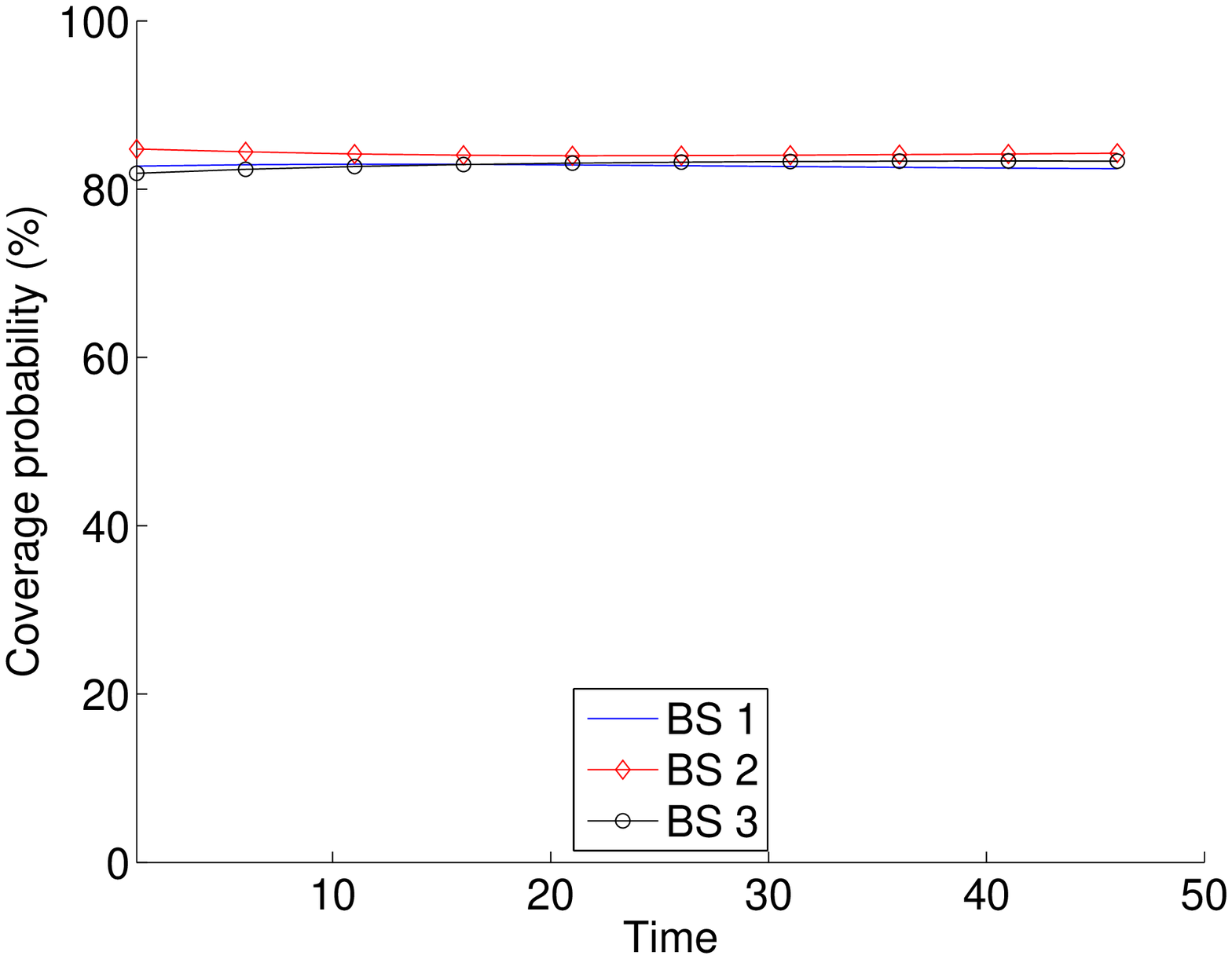}
		\end{center}
		\caption{Hexagonal network with coordination, coverage probability as a function of time}
		\label{fig:hexa_network_coverage_coord}
	\end{figure}

	We first consider an hexagonal network with $12$ \acp{BS} and inter-site distance of $500$ m, using a wrap-around to avoid border effects and ensure that all \acp{BS} are symmetric.  Namely the \acp{BS} are placed on a torus. We are interested in stability of the operating point $P^*$ , $P^*_i=46 dBm$ , $1 \leq i \leq I$. The corresponding target value for $G_i$ is $\overline{G}_i = 80\%$, $1 \leq i \leq I$. Without coordination, this operating point is unstable as shown by calculating the Jacobian of $G$ at $P^*$ using finite differences and calculating its eigenvalues. Another illustration of instability is given by plotting trajectories of the corresponding (non-linear) \ac{ODE} starting in a neighbourhood of $P^*$. On figures~\ref{fig:hexa_network_powers} and~\ref{fig:hexa_network_coverage} we represent the transmitted powers and coverage probabilities respectively as a function of time obtained by discretization of the \ac{ODE}. Only \acp{BS} $1$, $2$ and $3$ are represented for clarity , where \acp{BS} $2$ and $3$ are neighbors of \ac{BS} $1$. Clearly, $P^*$ is not stable and the plotted solution does not remain close to $P^*$. On figures~\ref{fig:hexa_network_powers_coord} and~\ref{fig:hexa_network_coverage_coord} we show the same quantities when the coordination mechanism is applied with $C = -(JG(P^*))^T$ as explained in the previous section. Indeed, the solution stays close to $P^*$ at all times. 

	Practical networks do not exactly follow an hexagonal model due to non-uniformity of traffic and scarcity of sites with good propagation characteristics. A popular model to take into account this irregularity is to assume that \acp{BS} locations follow a Poisson point process on the plane. Based on measurements from operational networks, results in \cite{AndrewsBaccelliGeometry} suggest that from the point of view of coverage, the Poisson model is pessimistic while the hexagonal model is optimist and reality lies somewhere in-between. We show that, like in the hexagonal case, instability occurs with a non-negligible probability. Hence instability is not an artifact of the hexagonal model. 
	
\begin{figure}[htbp]
		\begin{center}
			\includegraphics[width=\columnwidth]{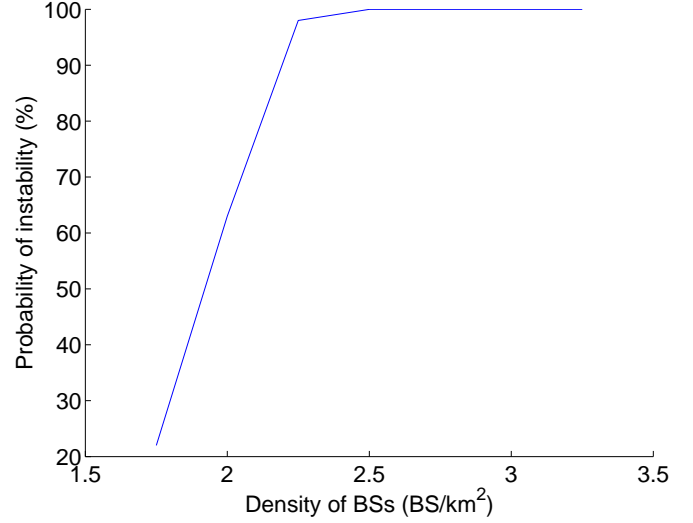}
		\end{center}
		\caption{Probability of observing an unstable network as a function of the density of BSs}
		\label{fig:poisson_network_proba}
	\end{figure}

	We use the following procedure. For each snapshot we generate \ac{BS} locations according to a Poisson process on a square area of $4 km^2$ , and we find a point $P^* = \Set{P_i^*}_{1 \leq i \leq I}$ at which all \acp{BS} have the same coverage probability. To be consistent with the hexagonal model, we choose the neighbors of a \ac{BS} to be the $6$ closest \acp{BS}. We calculate the Jacobian matrix of $G$ at $P^*$ to assess its stability. We simulate $100$ snapshots to estimate the probability of observing an unstable network. Figure~\ref{fig:poisson_network_proba} shows the probability of instability for different values of the number of \acp{BS} per unit of surface. There is a non-negligible probability of instability, and this probability rapidly goes to $1$ when the network becomes denser. An intuitive explanation is that when the network gets denser, \ac{BS} become closer to each other, so that the coupling between the corresponding \ac{SON} mechanisms becomes stronger and causes instability. 
%
%
	
\section{Conclusion}\label{sec:conclusion}
	In this paper we have studied the problem of coordinating multiple \ac{SON} entities operating in parallel.  Using tools from control theory and Lyapunov stability, we have proposed a coordination mechanism to stabilize the system. The mechanism can be implemented in a distributed fashion so it is scalable with respect to the number of \acp{SON}. We have shown that the mechanism remains valid in the presence of measurement noise, using stochastic approximation. Instability and coordination in the context of wireless networks have been illustrated with two examples. We have shown that even for two control loops, instability can occur, and the influence of network geometry has been investigated. An interesting continuation of this work would be to investigate coordination without linearity/linearization. Coordination for non-linear systems is more challenging because we cannot rely on local analysis and Lyapunov functions which are quadratic forms.
\bibliographystyle{IEEEtran}
\bibliography{IEEEabrv,main}
\appendices
\section{Asymptotic behavior of ODEs}\label{app:odes}
	Consider the \ac{ODE}:
\begin{equation}
	\dot{\theta} = F(\theta),
\end{equation}
	which we assume to have a unique solution for each initial condition defined on $\RR^+$. We write $\Phi(t,\theta(0))$ the value at $t$ of the solution for initial condition $\theta(0)$. We denote by $d_{{\cal U}}(\theta) = \infu{u \in {\cal U}} \norm{ \theta - u}$ the distance to set ${\cal U}$. We say that ${\cal U}$ is \emph{invarient} if $\theta(0) \in {\cal U}$ implies $\Phi(t,\theta(0)) \in {\cal U} $ , $t \in \RR^+$. We say that ${\cal U}$ is \emph{Lyapunov stable} if for all $\delta_1 > 0$ there exists $\delta_2 > 0$ such that $d_{{\cal U}}(\theta(0)) \leq \delta_2$ implies $d_{{\cal U}}(\Phi(t,\theta(0))) \leq \delta_1$ , $t \in \RR^+$. A compact invariant set ${\cal U}$ is an \emph{attractor} if there is an open invariant set ${\cal A}$ such that $\theta(0) \in {\cal A}$ implies $d_{{\cal U}}(\Phi(t,\theta(0))) \tends{t}{+\infty} 0 $ . ${\cal A}$ is called the \emph{basin of attraction}.  A compact invariant set ${\cal U}$ is \emph{locally asymptotically stable} if it is both Lyapunov stable and an attractor. If its basin of attraction ${\cal A}$ is equal to the whole space then ${\cal U}$ is \emph{globally asymptotically stable}. Asymptotic stability is often characterized using \emph{Lyapunov functions}. A positive, differentiable function $V:\RR^I \to \RR$, is said to be a Lyapunov function if  $t \mapsto V(\Phi(t,\theta(0)))$ is decreasing, and strictly decreasing whenever $V(\Phi(t,\theta(0))) > 0$. Then the set of zeros of $V$ is locally asymptotically stable. If we add the condition $V(\theta) \tends{\norm{\theta}}{+\infty} +\infty$, then we have global asymptotic stability.

\section{Linear ODEs}\label{app:linear_ode}
	Consider the \ac{ODE}:
\begin{equation}\label{eq:son_linear_ode_app}
	\dot{\theta} = A (\theta - \theta^*).
\end{equation}
Its solution has the form:
\begin{equation}\label{eq:son_linear_ode_sol}
	\theta(t) =  \theta^* +  e^{t A} (\theta(0) - \theta^*).
\end{equation}
	We denote by $\prec$ positive negativity for symmetric matrices. $\theta^*$ is asymptotically stable iif all the eigenvalues of $A$ have a strictly negative real part. Alternatively, asymptotic stability applies iif there exists $0 \prec X$ such that $A^T X +  X A \prec 0$. In this case, $V(\theta)= (\theta - \theta^*)^T  X (\theta - \theta^*)$ is a Lyapunov function for the \ac{ODE}. The reader can refer to \cite{LMIControlBoyd} for the linear matrix inequality approach to stability.
\end{document}